\documentclass[aps,prl,twocolumn,superscriptaddress,reprint,showkeys]{revtex4-2}
\usepackage{graphicx}
\usepackage{amssymb, amsmath}
\usepackage{booktabs}
\usepackage[usenames]{color}
\usepackage{bm}
\usepackage{multirow}
\usepackage{dcolumn}
\usepackage{hyperref}
\usepackage{enumerate}
\usepackage[mathlines]{lineno}
\usepackage{textcomp}
\hypersetup{
    colorlinks=true,
    linkcolor=blue,
    filecolor=gray,      
    urlcolor=blue,
    citecolor=blue,
}

\begin{document}

\title{Inverse Melting of Polar Order in Chemically Substituted BaTiO$_{3}$}

\newcommand{\harvard}{ The Rowland Institute at Harvard, Harvard University, Cambridge, MA, 02138, USA}
\newcommand{\inde}{ Independent Researcher, Seattle, WA, 98052, USA}
\newcommand{\rutger}{Rutgers Center for Emergent Materials and Department of Physics and Astronomy, Piscataway, NJ, 08854, USA}

\affiliation{\harvard}
\affiliation{\inde}
\affiliation{\rutger}

\author{Yang \surname{Zhang}}
\email{yzhang6@fas.harvard.edu}
\affiliation{\harvard}

\author{Suk Hyun \surname{Sung}}
\affiliation{\harvard}

\author{Colin \surname{B. Clement}}
\affiliation{\inde}

\author{Sang-Wook \surname{Cheong}}
\affiliation{\rutger}

\author{Ismail \surname{El Baggari}}
\email{ielbaggari@fas.harvard.edu}
\affiliation{\harvard}

\begin{abstract}
In many condensed matter systems, long range order emerges at low temperatures as thermal fluctuations subside.
In the presence of competing interactions or quenched disorder, however, some systems can show unusual configurations that become more disordered at low temperature, a rare phenomenon known as ``inverse melting''.
Here, we discover an inverse melting of the polar order in a ferroelectric oxide with quenched chemical disorder (BaTi$_{1-x}$Zr$_{x}$O$_{3}$) through direct atomic-scale visualization using in situ scanning transmission electron microscopy.
In contrast to the clean BaTiO$_{3}$ parent system in which long range order tracks lower temperatures, we observe in the doped system BaTi$_{1-x}$Zr$_{x}$O$_{3}$ that thermally driven fluctuations at high temperature give way to a more ordered state and then to a re-entrant disordered configuration at even lower temperature.
Such an inverse melting of the polar order is likely linked to the random field generated by Zr dopants, which modulates the energy landscape arising from the competition between thermal fluctuations and random field pinning potential.
These visualizations highlight a rich landscape of order and disorder in materials with quenched disorder, which may be key to understanding their advanced functionalities.
\end{abstract}


\maketitle

The concepts of order and disorder permeate a broad spectrum of physical phenomena, spanning condensed matter \cite{cheng2011atomic}, liquids \cite{finney1970random}, and living cells \cite{schrodinger1946life}.
Transitions from disorder to order can be described by an `order parameter' whose value changes from zero conventionally in the disordered phase to finite in the ordered phase \cite{landau1937theory}. 
Temperature plays a key role in this process:
disorder is generally associated with high temperatures because thermal fluctuations encourage the exploration of microstates, disrupting perfect arrangements and decreasing the order parameter. 
Because thermodynamic systems minimize their Free Energy, the energy cost of misalignment can be overcome as the system gains entropy. 
As temperature decreases, disordered configurations yield to increasingly aligned and low energy arrangements. 
A classic example is the transition from paramagnetism to ferromagnetism, where the magnetization, representing the sum of spins, serves as the order parameter \cite{toledano1987landau}. 
In the paramagnetic phase, spins are disordered and uncorrelated, resulting in an average magnetization of zero. 
Upon cooling, spins increasingly align with one another and lead to long range correlations and finite magnetization.

The picture of thermodynamic order and disorder, however, becomes blurry when real materials incorporate quenched or frozen disorder in their structure \cite{aizenman1989rounding, fisher1991thermal}.
These imperfections introduce random fields into the system that can pin local configurations and give rise to near-degeneracies or frustration, leading to rich phenomena like memory, hysteresis, and temperature chaos \cite{kaul1985static, dotsenko1995critical,grant1987metastable}. 
One such disorder-induced phenomenon is inverse melting, in which order gives way to a more disordered state at low temperature \cite{greer2000too}. 
Proposed a century ago \cite{tammannkristallisieren}, this unusual melting has been observed sparingly in systems such as polymers \cite{rastogi1999unusual},  metal alloys \cite{sinkler1997transmission}, magnetic thin films \cite{portmann2003inverse}, vortex lattices in superconductors \cite{avraham2001inverse}, and more recently in ferroelectric domains \cite{nahas2020inverse}. 
Various spin models realizing inverse melting show that the transition can be either first-order or continuous, depending on the parameters and nature of disorder  \cite{schupper2004spin,schupper2005inverse,das2023inverse}.
In all these cases, the inverse melting can be divided into two categories: changes in states of matter, like crystalline to amorphous transitions; or  variations of large mesoscale structures on the order of 0.1-10 $\mu m$, such as magnetic and ferroelectric domains.

An intuitive question arises: can inverse melting emerge from atomic- and nano-scale fluctuations of the local order instead of changes to mesocale domains or states of matter?
Spin glasses provide a realization of this scenario in magnetic systems, where the re-entrant spin glass transition can be achieved at lower temperature by balancing the spin interaction and internal frustration \cite{aeppli1983spin,dho2002reentrant}.
But in their ferroelectric counterparts, this phenomenon remains unexplored. 
The order parameter in a ferroelectric is polarization, arising from the off-centering of ions and the breaking of inversion symmetry \cite{cowley1980structural}. 
Typically, ferroelectricity requires the long-range alignment of polarization.
However, random fields created by quenched disorder may suppress the long-range order of polarization and broaden or blur phase transitions, leading to new ferroelectric phases with polar order fluctuating at nano- or atomic-scale, such as relaxor and superparaelectric states \cite{cross1987relaxor,shvartsman2012lead}.
Despite their broad appeal across fields \cite{kutnjak2006giant, pan2021ultrahigh}, how random fields regulate the local structure, spatial correlations, and detailed structure of disorder in ferroelectrics remain mostly inaccessible \cite{xu2006electric, krogstad2018relation, kumar2021atomic, eremenko2019local}, particularly how they evolve with temperature \cite{takenaka2017slush}.  
In this work, we studied a chemically doped ferroelectric oxide BaZr$_{0.2}$Ti$_{0.8}$O$_{3}$ (BZTO) which exhibits a rich phase diagram with different ferroelectric states.  
The random field in this system arises from chemical dopants that modify the local environment. 
With 20\% Zr doping, the BZTO exhibits novel ferroelectric states with so-called "diffuse phase transition", where the well-known ferroelectric state of BaTiO$_{3}$ (BTO) no longer appears.
This state is distinct from relaxors, as it does not show frequency dependent behavior \cite{shvartsman2012lead}. 
To reveal the fluctuation of microscopic order parameters, we employed aberration-corrected scanning transmission electron microscopy (STEM), which is powerful to capture the polar order at atomic-scale in ferroelectric oxides \cite{nelson2011spontaneous, tang2015observation, dong2019super}.
Through direct, atomic-scale mapping of the polar order and its nanoscale fluctuations across various temperatures, we discover that polar order in this doped oxide exhibits inverse melting at low temperature.

\begin{figure}
    \centering
    \includegraphics[width=\linewidth]{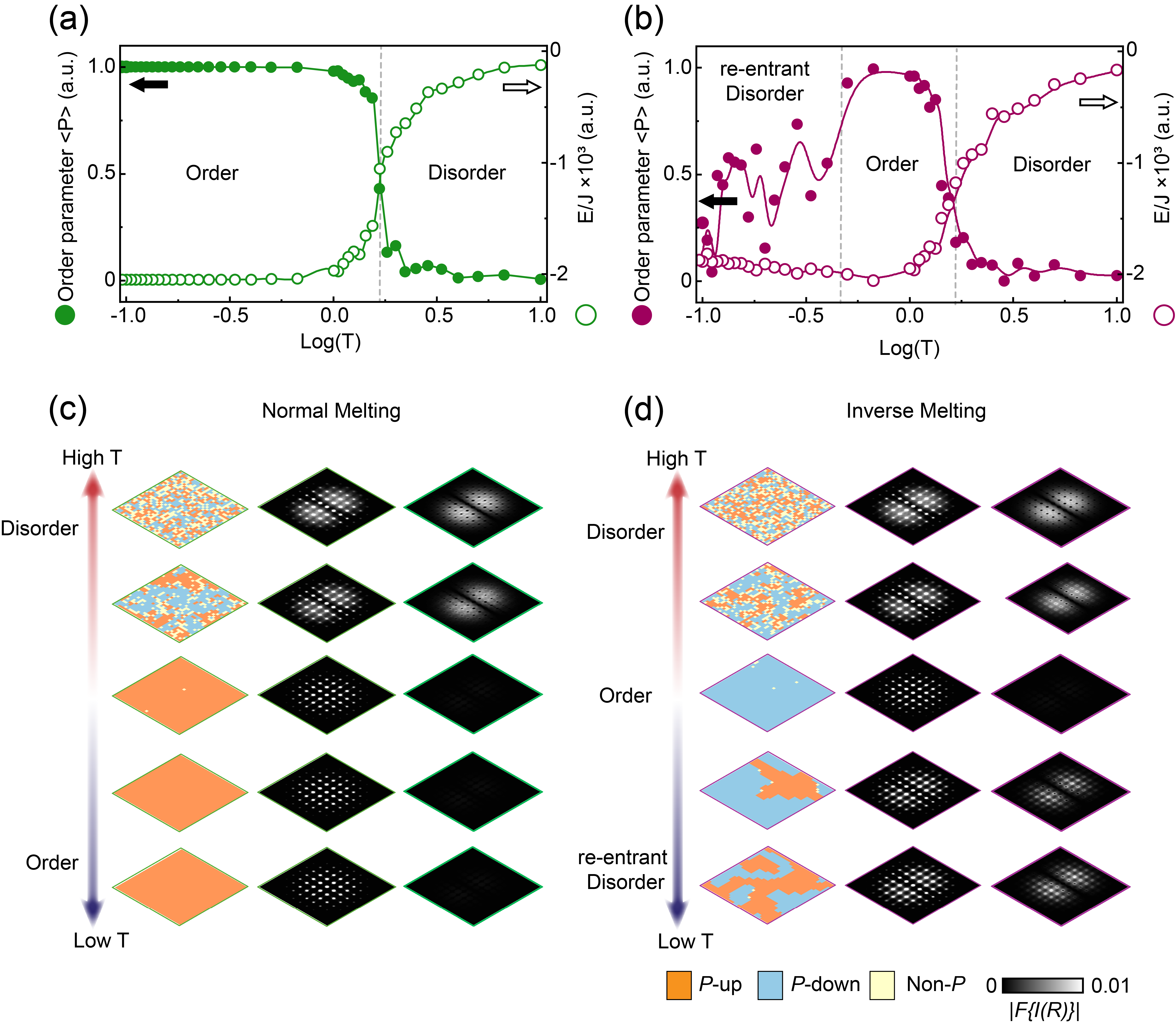}
    \caption{Inverse melting in the presence of a random field.
    (a, b) Evolution of the order parameter (solid circle) and energy (hollow circle) as a function of temperature in a two-dimensional spin-1 random field Ising model. 
    The (a) and (b) represents the evolution in the absence and presence of random field, respectively.
    (c, d) left panels: Snapshots of real-space configurations extracted at different temperatures in the absence (c) and presence (d) of a random field.
    The orange, blue and yellow represents that order parameter equal to 1 (polar up), -1 (polar down) and 0 (non polar), respectively.
    Middle panels: Fourier transformation of 2-dimensional pattern constructed from real-space configurations.
    Right panels: 2D difference plot after subtraction with Fourier transformation of ordered 2D pattern, which is used to highlight the diffuse intensity.
    The presence of diffuse intensity reflects the degree of disorder.
    }
    \label{fig1}
\end{figure} 

\begin{figure*}
    \centering
    \includegraphics[width=0.9\linewidth]{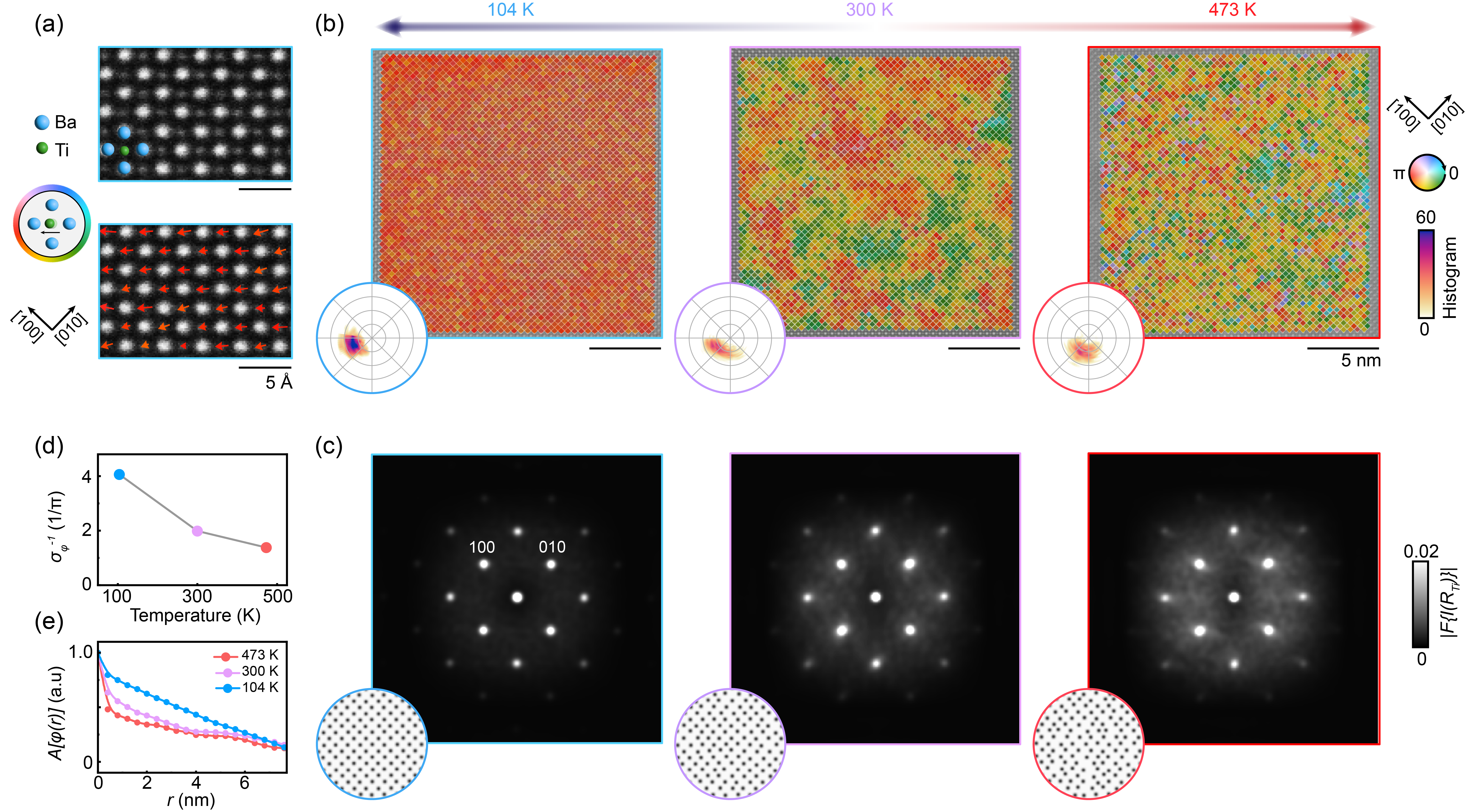}
    \caption{Atomic-scale visualizations of polar order across normal melting in BaTiO$_{3}$.
    (a) Upper panel: the ADF-STEM image of [001]-projected BaTiO${_3}$ collected at 104 K.
    The atomic model of Ba and Ti is overlaid on the image.  
    Lower panel: the measured polar displacement of Ti deviating from the central position determined by four neighboring Ba atoms
    The color represents the angle of polar displacement.
    (b) Real space maps of polar displacement angles at three temperatures, 473 K, 300 K and 104 K. 
    The color and transparency represents the direction and amplitude of polar displacement.
    The left corner is the polar histogram plot of polar displacements at the three temperatures.
    The radius is from 0 to 40 pm.
    (c) Fourier transform of the real space Ti positions at three temperatures.
    The left corner shows the cropped real space Ti positions (raw data is shown in Fig. S4a). 
    The amplitude of polar displacements is magnified 10 times to highlight the difference in positional disorder across the three temperatures.  
    (d) The polar displacement angle deviations ($\sigma_{\phi}^{-1}$).
    (e) Auto correlation function of polar displacement angle $A[\phi(\textit{r})]$ at three temperatures.}
    \label{fig2}
\end{figure*}

We first illustrate how a random field promotes degenerate energy landscapes and re-entrant disordered states at low temperature, using a simple two-dimensional spin-1 random field Ising model (see methods).
We define the order parameter, or polarization, at site $i$ as $P_{i}\in\{0,\pm 1\}$, where $P_{i}$ = 0 represents the non-polar phase and $P_{i}$ = $\pm 1$ two distinct polar states.
The Hamiltonian, $H$, is given by 
\begin{equation*}
  \mathcal{H} \displaystyle = -J\sum_{<ij>}{P_{i}P_{j}} - \sum_{i} h_{i}P_{i}  
\end{equation*}
where $J>0$ is the exchange energy, $h_{i}$ is a local random field normally distributed around zero and $<..>$ is the sum over nearest neighbors.
The system can lower its energy through the first term by aligning to a finite polarization. but the random field locally favors or disfavors a specific polarization state.
Figure 1a shows the temperature evolution of the order parameter and energy in the absence of a random field; 
as expected, the disordered state at high temperature gives way to a uniform long-range ordered state at low temperature.
Upon adding a random field, the thermodynamics of the order parameter are altered (Fig. S1).
Within a specific range, the system first enters a more ordered state at intermediate temperatures and then, a re-entrant disordered phase emerges as pinned regions and near-degeneracies break and reform at different temperatures.
Figure 1b shows this increase in the energy and a drop in the average polarization with decreasing temperature.
The left panels of Figs. 1c and d show representative real-space snapshots of the order parameter at different temperatures without and with the random field, highlighting the emergence of inverse melting in the latter case.
These two different melting behaviors can also be recognized by the change of diffuse intensity in reciprocal space.
Disorder contributes diffusion around the sharp peak, as shown in the right panels of Figs. 1c and d.  
Such an evolution is consistent with the presence of a rough energy landscape and frustrated, highly degenerate, metastable configurations induced by the random fields.

Using atomic-resolution imaging of polar order, we explore the evolution of the local ordering across temperature without and with a random field.
We first exemplify such capability through atomic-scale visualizations of pure BTO.
Pure BTO is an archetypal ferroelectric oxide, whose order parameter relates to the off-center displacement of Ti atoms relative to the neighboring O atoms \cite{cochran1960crystal}.
Such displacements can be mapped quantitatively for each unit cell by fitting atomic positions of Ba and Ti in annular dark-field (ADF-STEM) data, a capability extended to both cryogenic and high temperature. 
The upper panel of Fig. 2a shows an ADF-STEM image of BTO at 104 K, with the Ba atoms appearing brighter than Ti atoms due to their larger atomic number.
The lower panel shows a ADF-STEM image overlaid with the arrows representing Ti displacements and the color representing the direction of the displacements.
The displacements are oriented towards the projection of the $\langle 111\rangle$ direction, consistent with the rhombohedral phase of BTO at this temperature \cite{megaw1947temperature}.

Figure 2b shows the evolution of the real space maps of polar displacements at 104 K, 300 K and 473 K.
In contrast to the uniform ordered state at low temperature, domains with spatially fluctuating polar angles appear at room temperature (see SI notes for more details).
The spatial fluctuations of polar order increases significantly at higher temperature, with the transition to local polar order (see SI notes for more details), although the size of domains is usually underestimated due to projection nature of STEM image.
The influence of oxygen vacancy generated at high temperature is excluded (Fig. S8).
Additionally, the polar histogram shown in the insets exhibit the average amplitude of polar displacement increased from 10 pm to 14 pm and to 19 pm with lower temperatures. 
This melting from an ordered to disordered state upon heating is further confirmed through the diffuse intensity in the Fourier transforms.
While the sharp Bragg peaks encode the primary crystalline periodicity of the lattice, diffuse intensity encodes disorder and fluctuations in the positions of the atomic positions (Fig. S3). 
Figure 2c shows a Fourier transform of the real space Ti positions (inset of Fig. 2c and Fig. S4a).
With higher temperature, this diffuse intensity between Bragg peaks becomes more prominent, consistent with evolution shown in Fig. 1c.

To quantify the degree of order ferroelectric displacements, we extract polar displacement angles, $\phi(\mathbf{r})$ , their standard deviation, $\sigma_{\phi}$, and the correlation function, $A[\phi(\mathbf{r})]$.
Figures 2d depict $\sigma_{\phi}^{-1}$ measured from the statistical distribution of the polar order at three temperatures (inset of Fig. 2b) . 
The increase in $\sigma_{\phi}^{-1}$ at 104 K is consistent with the ordered state that we visualize in the real space.
Figure 2e shows the profiles of the autocorrelation function extracted from [110] direction, $A[\phi(\mathbf{r})]$, whose decay encodes the correlation length of the polar order.
From 104 K to 473 K, the decay behavior of $A[\phi(\mathbf{r})]$ shows slower drop at lower temperature, as expected from thermally driven fluctuations. 
The evolution of the degree of polar order can therefore be directly visualized in ADF-STEM image and quantified by changes in $\sigma_{\phi}^{-1}$ and decay in $\phi(\mathbf{r})$ correlations.

\begin{figure}
    \centering
    \includegraphics[width=\linewidth]{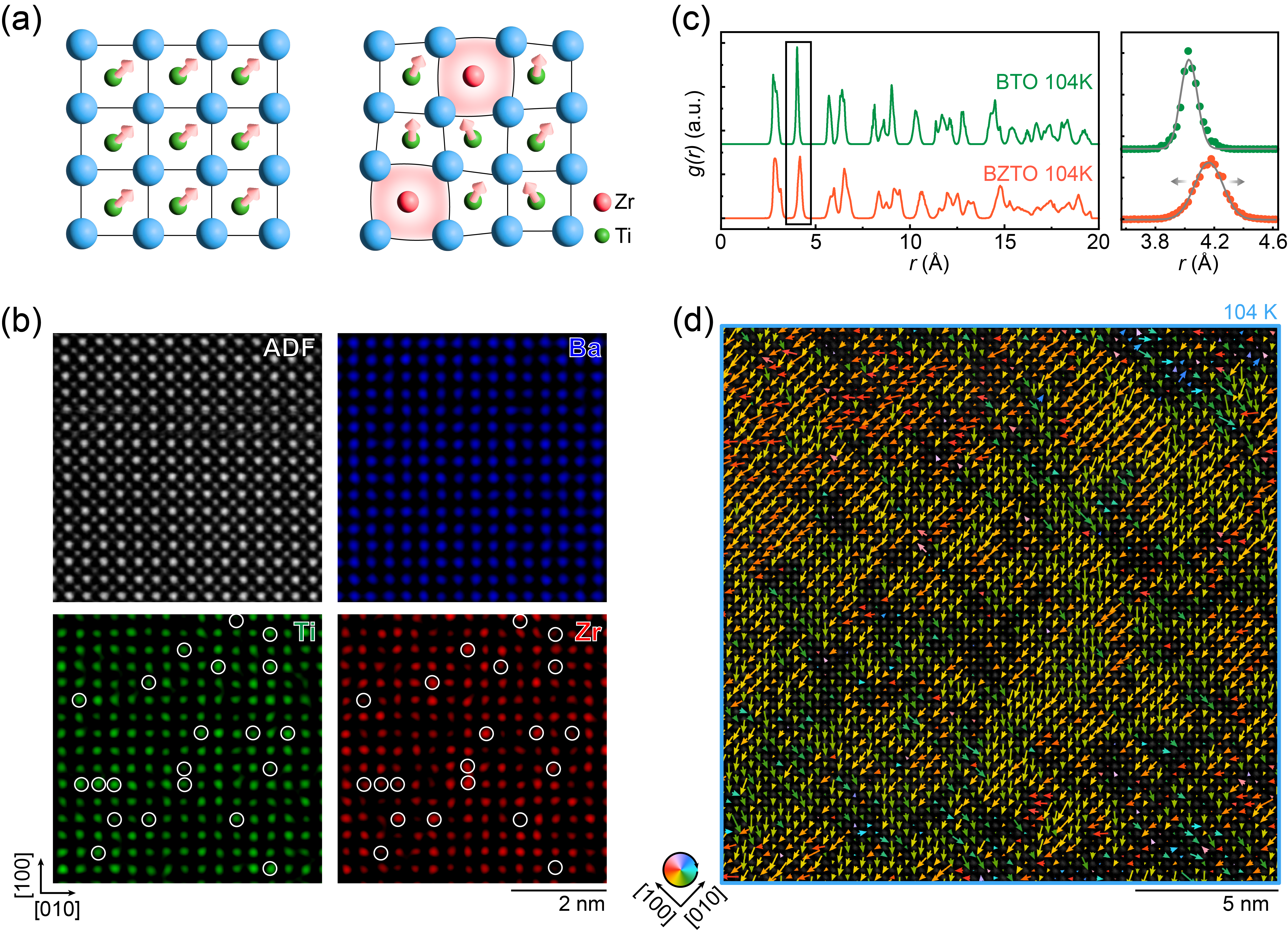}
    \caption{Atomic-scale mapping of chemically quenched disorder and accompanied random field in BaZr${_{0.2}}$Ti${_{0.8}}$O${_3}$.
    (a) The schematic shows the generation of random field by Zr-doping. 
    The Zr dopant locally expands the lattice and disturbs ferroelectric order.
    (b) Atomic-resolution EDS chemical mapping of Ba, Zr, and Ti signals collected at 300 K. 
    The white circles highlight representative positions with negative correlation between Zr and Ti.
    (c) Radial pair distribution function $g(\textit{r})$ of BaTiO${_3}$ and BaZr${_{0.2}}$Ti${_{0.8}}$O${_3}$ at 104 K. 
    The right panel shows an enlarged view of the next-nearest atomic spacing. 
    The gray line shows a Gaussian fit to the peaks. 
    The arrow helps distinguish the broaden of the peak in Ba(Ti, Zr)O${_3}$. 
    (d) ADF-STEM image overlaid with polar displacements in BaZr${_{0.2}}$Ti${_{0.8}}$O${_3}$ at 104 K. 
    The projection is along the [001] zone axis. 
    The scale bar is 5 nm.
     }
    \label{fig3}
\end{figure}

The random field in real materials can have multiple characters such as electrostatic (due to charge disorder), magnetic (due to magnetic impurities), or elastic (due to local strain). 
In BTO, isovalent dopants, such as Sn$^{4+}$, Zr$^{4+}$, do not introduce significant charge disorder and have been found to be an effective mean to tune polar order without the complications of charge inhomogeneity \cite{hennings1982diffuse}.
The Zr dopants will alter polar order of BZTO in two ways (Fig. 3a):
(i) The mismatch between the Zr dopants and the pristine Ti will create an internal strain field around the doping sites.
(ii) the Zr dopants is polar inactive \cite{mentzer2019phase}, which breaks the local alignment of ferroelectric correlation between polar active Ti ions.
In this way, if there exists the atomic-scale fluctuations of the Zr dopants, it will generate a random field in the material.
Using atomic-resolution energy-dispersive dispersion spectroscopy (EDS), we visualize the distribution of Zr dopants in BZTO.
Figure 3b reveals fluctuations in Zr and Ti signals, and a homogeneous Ba signal, ruling out chemical segregation of Zr dopants.
The effect of Ga injection during sample preparation is excluded (Fig. S9). 
Moreover, there exists a negative correlation between Zr and Ti, with some representative sites highlighted by white circles. 

\begin{figure}
 \centering
    \includegraphics[width=\linewidth]{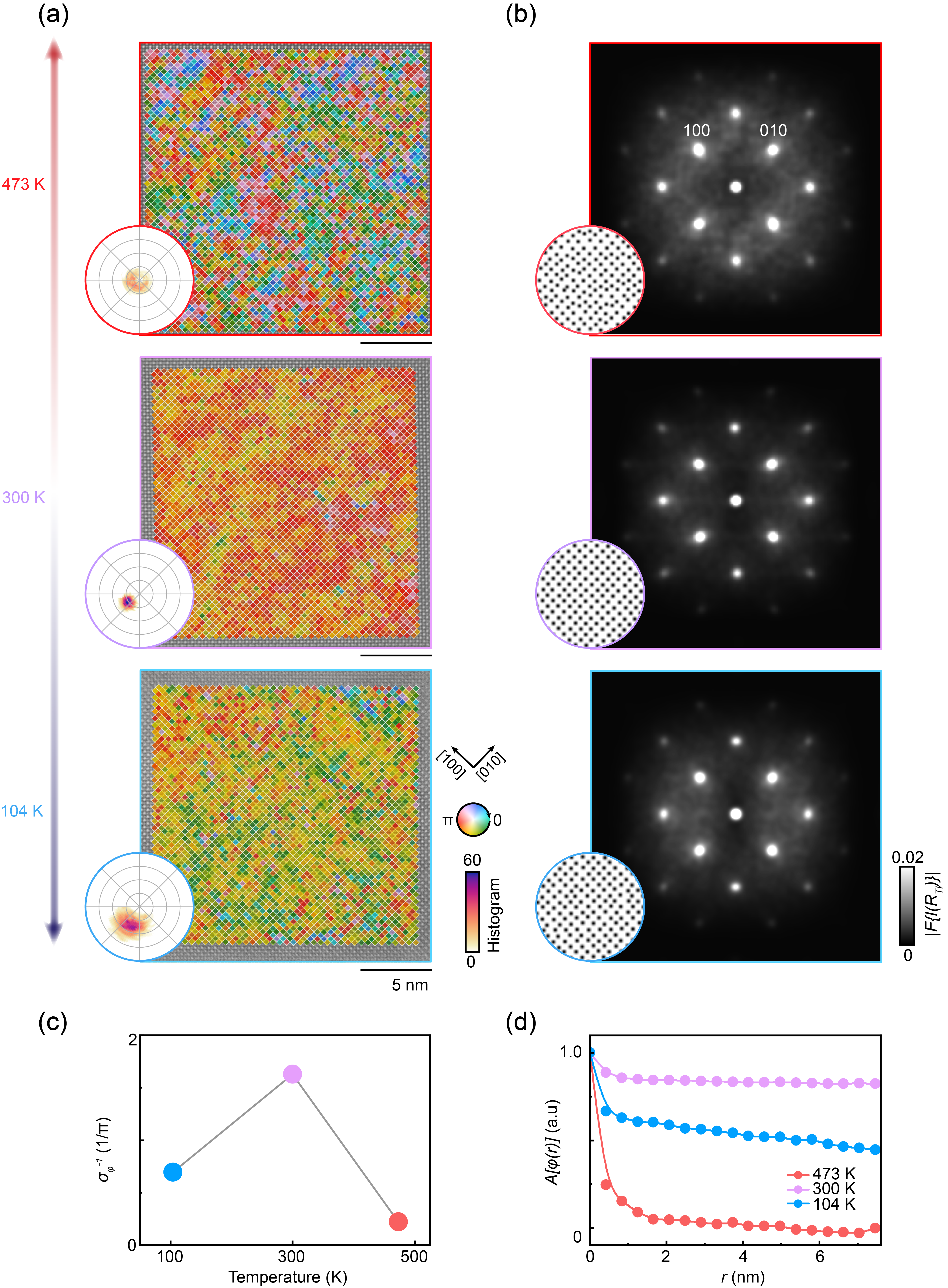}
    \caption{Inverse melting of polar order in BaZr${_{0.2}}$Ti${_{0.8}}$O${_3}$.
    (a) Real space maps of polar displacement angles at three temperatures, 473 K, 300 K and 104 K. 
    The color and transparency represents the direction and amplitude of polar displacement.
    The left corner is polar histogram plot of polar displacement at three temperatures. 
    The radius is from 0 to 40 pm.
    (b) Fourier transform of the real space Ti positions with normalized displacement at three temperatures.
    The left corner shows the cropped real space Ti positions (raw data in Fig. S4b). 
    The amplitude of polar displacement is magnified 10 times to highlight the difference among three temperatures. 
    (c) The polar displacement angle deviation ($\sigma_{\phi}^{-1}$) measured at three temperatures.
    (d) Auto correlation function of polar displacement angle $A[\phi(\textit{r})]$ at three temperatures.}
    \label{fig4}
\end{figure}

To confirm the random field into the system, we then compare the internal strain field and polar order in the presence and absence of Zr dopants.
Measurements of the relation between lattice constant and Zr occupation show that BZTO exhibits an evidently positive correlation (Figs. S9-S10), confirming the role of Zr dopants on creating an internal strain field.
The pair distribution $g(\textit{r})$ can be employed to compare the inhomogeneity of strain field. 
By taking the atomic positions fits in the ADF-STEM data, we calculated $g(\textit{r})$ of BTO and BZTO at 104 K, respectively (see methods for more details).
As shown in Fig. 3c, the $g(\textit{r})$ peaks of BZTO appear broader than those of BTO at 104 K, indicating the existence of an inhomogeneous strain fields and positional disorder in BZTO.  
Furthermore, the polar order in BZTO at 104 K (Fig. 3d) shows considerably more spatial fluctuations at the nanoscale when compared to the undoped system.
This highlights that fluctuation of Zr dopants does indeed create a random field to induce inhomogeneous internal strain and disturb the alignment of ferroelectric polarization.

Remarkably, the random field generated by Zr dopants creates an inverse melting, whereby the low-temperature state is more disordered than the room-temperature phase. 
Figure 4a displays the real space distribution of polar displacement at the three representative temperatures.
At 473 K, polar displacements are still measurable (see SI notes for more details), with averaged amplitude close to 7 pm and highly disordered due to the dominance of thermal fluctuations.
Cooling to room temperature, polar displacements begin to order and averaged amplitude increased to 11 pm.
Upon further cooling to 100 K, the averaged amplitude keeps increasing to 17 pm, but the system remains disordered instead of forming a uniform state, in marked contrast to the undoped BTO system.
These observations are further supported by the diffuse intensity in the Fourier transforms of atomic positions (Fig. 4b), which strongly resembles the behavior shown in Fig. 1d.
This diffuse intensity in the Fourier transform represents disorder in the polar displacement instead of size effects (see SI notes for more details).
Additionally, by extracting $\sigma_{\phi}^{-1}$ and $A[\phi(\mathbf{r})]$, we observe that the 100 K phase is, in fact, more disordered than the 300 K phase (Figs. 4c-d).
The variation of $\sigma_{\phi}^{-1}$ is larger than the measured precision estimated by BaTiO$_{3}$ (Fig. S15).
The $A[\phi(\mathbf{r})]$ shows non-monotonic change, first increasing from 473 K to 300 K and then dropping at 104 K.
Data collected across different regions/domains shows the same evolution (Fig. S16).
This indicates an inverse melting involving re-entrant disordering of polar order upon cooling.
This abnormal dynamic behavior is different from the evolution of competing order in relaxors (Fig. S17).

Our ability to map and analyze the correlations of polar order across temperature proved instrumental in unraveling the inverse melting phenomenon in a ferroelectric at the atomic scale.
Inspired by the visualizations, we presented a simple mapping between BZTO and a spin-1 random-field Ising model. 
Despite its simplicity, this model shows that a random field, which can be generated by the presence of 
fluctuated Zr dopants, mediates the following sequence of transitions:
(i) at high temperature, large thermal fluctuations dominate the disordering process regardless of the strength of the random field; 
(ii) at intermediate temperature, thermal fluctuations are reduced but still significant enough to overcome and smear the random field potential, and so the system takes advantage of the exchange energy by aligning polarization; 
(iii) finally, at much lower temperature, the system is strongly pinned to the local random field, and so the local correlations are dominated by the coupling to the quenched disorder, leading to more disordered configurations than at higher temperature.
The role of Zr dopants suggests that this phenomenon might prove more general and applicable to other chemically doped ferroelectric systems, when the random field is strong enough to pin the polar order at low temperature. 
In addition to these isovalent dopants, heterovalent dopants will further induce an electrostatic random field, potentially driving more complex melting behavior of polar order in ferroelectric systems.
Furthermore, providing more theoretical insights into the relation between doping level and inverse melting is also compelling. 
A series of follow-up studies into lightly doped ferroelectrics and polar order at low temperature are required. 

Beyond the intriguing fundamental aspects, inverse melting holds appeal for its potential for practical applications. 
Thermodynamic properties, such as thermal expansion and electrocaloric effects, are intimately linked to changes in entropy within a system \cite{braun2018charge,li2022seawater,qian2014giant}. 
A state of increased disorder at low temperature could theoretically exhibit negative thermal expansion \cite{liu2014thermal, wendt2019entropic,lohaus2023thermodynamic}, which plays a key role in tailoring the overall thermal expansion coefficient of materials \cite{chen2008zero, chen2011role}. 
$g(\textit{r})$ extracted from atomic-resolution data show that while the BTO lattice contracts as polar order from 300 K to 104 K, the BZTO lattice expands in the low temperature disorder phase (Fig. S18). 
This behavior suggests that inverse melting and random fields may be unexplored, generic mechanisms for generating negative thermal expansion.
These atomic-scale visualizations open new avenues for probing and designing material properties by means of controlled disorder and correlations.

\section{Conflict of interest}
The authors declare no conflict of interest.

\section{Acknowledgments}
This work was supported by the Rowland Institute at Harvard.
Focused ion beam sample preparation was performed at the Harvard University Center for Nanoscale Systems (CNS); a member of the National Nanotechnology Coordinated Infrastructure Network (NNCI), which is supported by the National Science Foundation under NSF award no. ECCS-2025158. 
Transmission electron microscopy was carried out through the use of MIT.nano's facilities.
S.-W. C. was supported by the W. M. Keck foundation grant to the Keck Center for Quantum Magnetism at Rutgers University.
We thank Andrew Murray, Rosy Hosking and Robert Hovden for feedback during manuscript preparation.

\section{Author contributions}
Y. Z. conceived the study. 
S.-W. C provided bulk samples.
Y. Z. prepared the in situ TEM samples, performed in situ electron microscopy experiments and analyzed data with help from S. H. S. and I. E.. 
I. E., C. B. C. carried out theoretical calculations.
Y. Z. and I. E. wrote the paper. 
All authors discussed the results and commented on the manuscript.

\bibliographystyle{ref}
\bibliography{reference}

\end{document}